\documentclass{elsart3}

\usepackage{amssymb}
\usepackage{graphicx}

\begin{document}

\begin{frontmatter}



\title{Pressure-induced valence change in the rare earth metals:\\The case of Praseodymium\thanksref{label1}}
\thanks[label1]{This paper was presented in the international conference Rare Earth '04 in Nara Japan.It will appear in Journal of Alloys and Compounds.}

\author[a,b]{Naoyuki Tateiwa}
\author[b]{Akitoshi Nakagawa}
\author[b]{Kazuhiko Fujio}
\author[b]{Tatsuya Kawae}
\author[b,c]{Kazuyoshi Takeda}

\address[a]{ Advanced Science Research Center, Japan Atomic Energy Research Institute, Tokai, Ibaraki 319-1195, Japan}
\address[b]{Department of Applied Quantum Physics, Faculty of Engineering, Kyushu University, Hakozaki, Fukuoka 812-8581, Japan}
\address[c]{Institute of Environmental Systems, Faculty of Engineering, Kyushu University, Hakozaki, Fukuoka 812-8581, Japan}

\begin{abstract}
  The rare earth metal praseodymium (Pr) transforms from the d-fcc crystal structure (Pr-III) to {$\alpha$}-U one (Pr-IV) at  20 GPa with a large volume collapse (${\rm\Delta} V/V$ = 0.16), which is associated with the valence change of the Pr ion. The two 4{\it f} electrons in the Pr ion is supposed to be itinerant in the Pr-IV phase. In order to investigate the electronic state of the phase IV, we performed the high pressure electrical resistance measurement using the diamond anvil cell up to 32 GPa. In the Pr-IV phase, the temperature dependence of the resistance shows an upward negative curvature, which is similar to the itinerant 5{\it f} electron system in actinide metals and compounds.  This suggests the narrow quasiparticle band of the 4{\it f} electrons near the Fermi energy.  A new phase boundary is found at $T_{0}$ in the Pr-IV phase. 
From the temperature and magnetic field dependences of the resistance at 26 GPa, the ground state of the Pr-IV phase is suggested to be magnetic. Several possibilities for the origin of $T_{0}$ are discussed.
\end{abstract}

\end{frontmatter}

\section{Introduction}
 Cerium (Ce), praseodymium (Pr) and gadolinium (Gd) show the pressure induced phase transition with a large volume collapse (${\rm\Delta} V/V$ $\sim$ 0.1) at 0.7, 20, and 60 GPa, respectively~\cite{rf:grosshans,rf:mcmahan1,rf:hua}. The case in Ce is well-known as the $\alpha$-$\gamma$ transition which has been studied for many years from both theoretical and experimental points of view~\cite{rf:soderlind1}. It is basically accepted that the transition at 0.7 GPa is associated with the drastic change of the electronic state in the Ce ion and that the 4{\it f} electron changes from the localized and itinerant states across the transition~\cite{rf:soderlind1}. It is also suggested from  theoretical and experimental studies that the transitions in Pr and Gd are associated with the drastic change of the 4{\it f} electrons state and that the 4{\it f} electrons becomes itinerant at the higher pressure region~\cite{rf:grosshans,rf:mcmahan1}. However contrary to the case in Ce, there are few experimental investigations on the transitions in two elements expect by the crystal analyses. 

 Figure 1 shows the pressure phase diagram of Pr based on Ref. 5~\cite{rf:zao}.  At room temperature, Pr shows crystal phase transitions from Pr-I (dhcp) to Pr-II (fcc) phases around 4 GPa, from Pr-II to Pr-III (d-fcc, $R{\bar{3}}m$) phases around 7 GPa, and Pr-III to Pr-IV ({$\alpha$}-U, $Cmcm$) phases at 20 GPa~\cite{rf:zao,rf:mao,rf:baer}. The transition between the Pr-III and IV phases is accompanied with a large volume collapse (${\rm\Delta} V/V$ = 0.16 )~\cite{rf:baer}. A simple thermodynamic consideration indicates the disappearance of the degree of freedom in the electronic state of the localized the 4{\it f} electrons in the Pr-IV phase~\cite{rf:baer}. The sequence of the crystal structure and the change in the volume across the transition are explained by theoretical studies which treated the 4{\it f} electrons in the Pr-IV phase as being itinerant~\cite{rf:svane,rf:soderlind2}.
 \begin{figure}[<t>]
\begin{center}
\includegraphics[width=7.5cm]{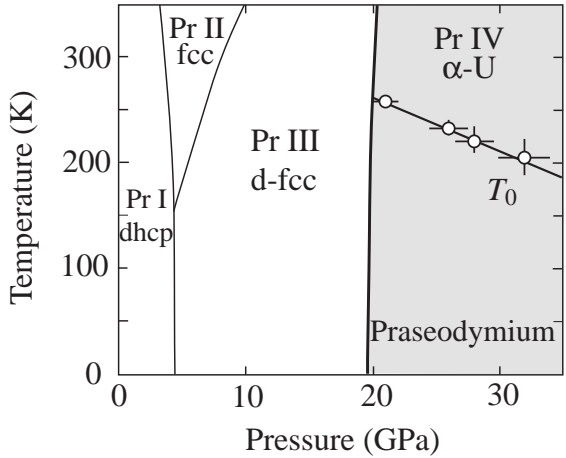}
 \end{center}
\caption{\label{fig:epsart} Pressure phase diagram based on Ref. 5. The bold line indicates the first order phase boundary with a large volume collapse (${\rm\Delta} V/V$ = 0.16 ) between the Pr-III and IV phases. $T_{0}$ is the new phase boundary which is found in this present study.}
\end{figure} 
The high pressure electrical resistance measurement on the Pr-II and III phases was preformed by Heinrichs and Wittig~\cite{rf:heinrichs}. The resistance shows a monotonous temperature dependence and there is no anomaly associated with the magnetic ordering or the superconductivity in the temperature region investigated (4.2-300 K). There is no report on the physical property of the Pr-IV phase expect the crystal analyses. In this paper, we report our experimental result of the electrical resistance measurement on Pr up to 32 GPa. 
 
\section{Experimental methods}
 We used a clamp-type diamond anvil cell (DAC) made from a non-magnetic Cu-Be alloy. The details of experimental procedures are in Ref. 11~\cite{rf:amaya}. Diamond anvils with the culet size of 0.5 mm were used. A sample was placed on the aluminum oxide (Al$_2$O$_3$) insulating layer on the non magnetic stainless steel (SUS310S) gasket. The insulating layer contained tiny amounts of epoxy which acted as a pressure transmitting medium. Four platinum-film electrodes with the thickness of 5 ${\mu}$m were placed and the distance between electrodes for the voltage measurement was within 0.10 mm in order to ensure the homogeneity of the pressure between electrodes. We used the poly crystal sample of Pr. The purity was 99.9\%. It was further annealed for 100 hours at 1000 $^{\circ}$C for purification. The electrical resistance measurement was carried out by the a.c four-terminal method using a resistance bridge (Linear Research, LR-700).  The resistivity (${\Omega}{\rm cm}$) of the sample was not obtained within present experimental configuration. We show the resistance (${\Omega}$) of the sample. Ruby chips with the diameter about 10 ${\mu}$m were put around the center of the sample and the pressure was determined by a conventional ruby-fluorescence method at room temperature. The low temperature measurement was done using a $^4$He cryostat and $^3$He-$^4$He dilution refrigerator.
 
\section{Results and Discussions}

  The pressure dependence of the resistivity at 293 K (not shown here) is quantitatively consistent with Ref. 10~\cite{rf:heinrichs}. We show the temperature dependence of the resistance of the Pr-IV phase in figure 2. For the sake of clarity, the data are normalized by each value at 273 K and are shifted upwards by values in parentheses. The resistance decreases with decreasing temperature and the curve is convex upward. The resistance follows $T^{\rm 2}$-law in the low temperature region. We did not observe the superconducting phenomenon in the lowest temperature (0.13 and 0.20 K at 26 and 32 GPa, respectively). 
\begin{figure}[<t>]
\begin{center}
\includegraphics[width=7.5cm]{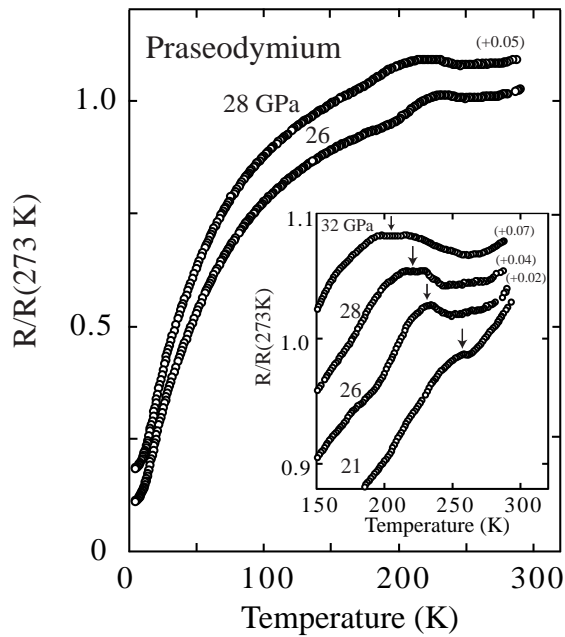}
 \end{center}
\caption{\label{fig:epsart} Temperature dependence of the resistance. Inset shows a high temperature part.}
\end{figure}

    We show the high temperature part of the resistance in the inset of the fig. 2. At the higher temperature side, there is a small hump in the resistance. The value of $T_{\rm 0}$ is defined as the peak position of the hump  and we show its pressure dependence in the pressure phase diagram in the Fig. 1. The value of $T_{\rm 0}$ decreases monotonously with increasing pressure. It is suggested that some kind of phase transition occurs at $T_0$. The anomaly at $T_0$ becomes more clearer with increasing pressure in the Pr-IV phase. It is suggested that the broaden hump structure at 28 and 32 GPa is due to the increment of the pressure distribution (${\rm\Delta} P$). The anomaly at $T_{\rm 0}$ would be sharp if the pressure environment were ideal. 
 
  The temperature dependence of the resistance  in the Pr-IV shows an upward curvature below 150 K and the resistance shows a tendency to saturate at higher temperature. This characteristic feature is similar to those observed in some actinide metals and compounds\cite{rf:brodsky}. This was explained by the effect of the narrow energy band near the Fermi energy from the theoretical point of views\cite{rf:jullien,rf:takaoka}. The present result suggests the narrow quasiparticle band of the 4{\it f} electrons near the Fermi energy in the Pr-IV phase. Comparison of the resistance curve with the theoretical model\cite{rf:jullien}, the width of the energy band near Fermi energy is roughly estimated to be the order of 100 K ($\sim$ 0.01eV) in the Pr-IV phase.  

    In the case of Ce,  the resistivity of the high pressure $\alpha$ phase shows a downward curvature\cite{rf:leger,rf:brodsky2}. From the photoemission experiment, the width of the energy band of the itinerant 4{\it f} electrons is estimated to be roughly the order of 1000 K ($\sim$ 0.1 eV) in the $\alpha$ phase\cite{rf:weschke}. This value is about ten times larger than that in the Pr-IV phase. 
\begin{figure}[<t>]
\begin{center}
\includegraphics[width=7.5cm]{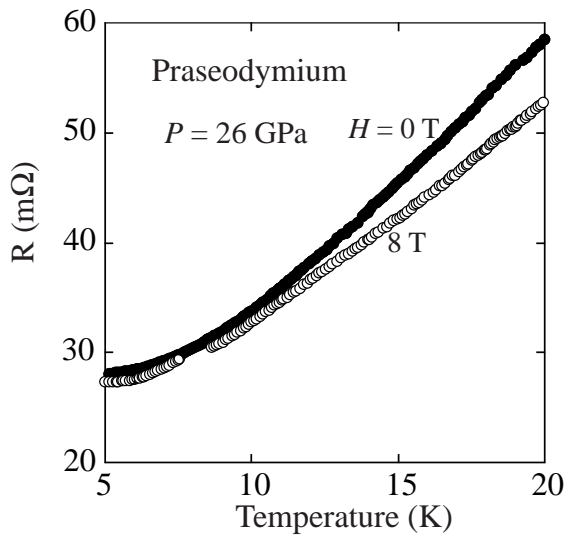}
 \end{center}
\caption{\label{fig:epsart} Temperature dependence of the resistance at 26 GPa under magnetic fields of 0 and 8 T.}
\end{figure}
      
  The temperature dependences of the resistance under magnetic fields of 0 and 8 T are shown in figure 3. The resistance at the higher temperature region decreases with increasing magnetic field. The absolute value of the negative magnetoresistance, -$[R(8 {\rm T})-R(0 {\rm T})]/R(0 {\rm T})$, is 2, 7, and 10\% at 4.2, 15, and 20 K, respectively. 

  The resistivity of a single crystal sample in usual metals shows the positive magnetic field dependence which reflects the topological character of Fermi surfaces~\cite{rf:ziman}. Also when there is a narrow 4{\it f} energy band near the Fermi energy, the negative magnetoresistance is observed if the magnitude of magnetic field becomes comparable with the band width. These  effects are the response of the ground state of the electronic state to magnetic field and thus become more distinct in the lower temperature region. This is obviously inconsistent with the present result in the fig. 3 where the magnetic field has a stronger effect at the higher temperature region~\cite{rf:comment}. 

 The present magnetic field dependence of resistance might reflect the response of the low energy magnetic excitation of the electronic state to magnetic field. A similar kind of magnetic field dependence in the resistance was observed in the magnetic ordered ground state of the itinerant 3{\it d} ferromagnet Sc$_3$In~\cite{rf:masuda}. The absolute value of the magnetoresistance -[$R(H)-R(0 {\rm T})$]  increases with increasing temperature and shows a strong enhancement around the transition temperature. This can be understood as follows. The resistance at finite temperature consists of the phonon part $\rho_{ph}$ and the magnetic part $\rho_{mag}$. The latter in the magnetic ordered state is originated from the scattering of the conduction electrons by the low energy magnetic excitations such as the magnon or spin fluctuation which are enhanced with increasing temperature. The magnetic field suppresses the scattering by the magnetic excitations. Therefore, -[$R({\rm H})-R(0 {\rm T})$] increases with increasing temperature if there is a very small contribution of the positive magnetoresistance as mentioned above. It should be noted that the magnetoresistance of Sc$_3$In were quantitatively explained on the frame work of the self-consistent renormalization theory of spin fluctuations~\cite{rf:moriya,rf:hioki}. The present result in the Fig. 3 suggests the magnetic ground state of the Pr-IV phase. There is no report of the crystal phase transition around $T_{\rm 0}$. Therefore, it is possible to speculate that $T_{\rm 0}$ is the magnetic ordering temperature of the itinerant 4{\it f} electrons. In order to check this speculation, the magnetization measurement is desirable but is almost impossible in such a high pressure region within the present technology of high pressure scientists. However, it might be possible to detect the internal magnetic molecular field by other methods such as the M$\ddot{\rm o}$ssbauer experiment where a measurement is available above 10 GPa recently. Meanwhile, at present stage, it is impossible to rule out completely the possibility of a crystal phase transition at $T_{0}$. It is also needed to re-investigate the change of the crystal structure around  $T_{\rm 0}$ in detail.

  \section{Conclusion}
 In conclusion, we performed the high pressure resistance measurement on Pr up to 32 GPa. The temperature dependence of the resistance in the Pr-IV phase shows an upward negative curvature. This suggests the narrow energy band of the itinerant 4 {\it f\,} electrons near the Fermi energy. The new phase boundary at $T_{0}$ is found and the pressure dependence of $T_{0}$ is obtained. Several possibilities of the origin of $T_{0}$ are discussed.
 



\begin{thebibliography}{00}
\bibitem{rf:comment} There is a break in the resistance at 8 T around 8 K. This is due to the miss-operation of a wiring system.
\end{thebibliography}
\end{document}